# Improvement of the critical temperature of NbTiN films on III-nitride substrates

Houssaine Machhadani[1], Julien Zichi[2], Catherine Bougerol[3], Stéphane Lequien[4], Jean-Luc Thomassin[1], Nicolas Mollard[4], Anna Mukhtarova[1], Val Zwiller[2], Jean-Michel Gérard[1], and Eva Monroy[1,*]

[1] Univ. Grenoble-Alpes, CEA-INAC-PHELIQS, 17 av. des Martyrs, 38000 Grenoble, France.
[2] KTH Stockholm, Department of Applied Physics, SE-114, 128 Stockholm, Sweden.
[3] Univ. Grenoble-Alpes, CNRS-Institut Néel, 25 av. des Martyrs, 38000 Grenoble, France.
[4] Univ. Grenoble-Alpes, CEA-INAC-MEM, 17 av. des Martyrs, 38000 Grenoble, France.
* eva.monroy@cea.fr

OrcIDs:
Eva Monroy: 0000-0001-5481-3267
Jean-Michel Gérard: 0000-0002-0929-0409

**Abstract**

In this paper, we study the impact of using III-nitride semiconductors (GaN, AlN) as substrates for ultrathin (11 nm) superconducting films of NbTiN deposited by reactive magnetron sputtering. The resulting NbTiN layers are (111)-oriented, fully relaxed, and they keep an epitaxial relation with the substrate. The higher critical superconducting temperature ($T_c$ = 11.8 K) was obtained on AlN-on-sapphire, which was the substrate with smaller lattice mismatch with NbTiN. We attribute this improvement to a reduction of the NbTiN roughness, which appears associated to the relaxation of the lattice misfit with the substrate. On AlN-on-sapphire, superconducting nanowire single photon detectors (SNSPDs) were fabricated and tested, obtaining external quantum efficiencies that are in excellent agreement with theoretical calculations.

# 1. Introduction

Superconducting ultrathin (few-nm-size) films are the keystone of devices such as superconducting hot electron bolometer mixers [1–4] or superconducting nanowire single-photon detectors (SNSPDs) [5–9]. Furthermore, the integration of such superconducting layers in semiconductor structures has been explored as an approach to improve the performance and add new features to high electron mobility transistors [10], and could grant access to interesting physical phenomena involving Majorana fermions [11] and topological insulators [12]. Niobium titanium nitride (NbTiN) is widely used in applications requiring nm-size films due to its physical and chemical stability. In comparison with NbN, NbTiN devices have demonstrated higher optical quantum efficiency, smaller recovery time, and reduced noise [13,14]. The origin of this improvement remains partially unclear, though it is generally attributed to a better crystalline quality, lower atomic concentration of oxygen, and higher metallic conductivity, which results in lower kinetic inductance [13,15].

Deposition of crystalline Nb(Ti)N materials by molecular beam epitaxy [10] or chemical vapor deposition [16] require relatively high substrate temperatures ($\approx$ 800°C and 900-1300°C, respectively). Therefore, compatibility with device processing imposes the deposition by sputtering, where polycrystalline ultrathin films with excellent morphology can be obtained in the temperature range of 25-300°C [17–19]. However, the superconducting performance of NbTiN layers, namely the critical temperature at which the phase transition from resistive to superconducting state occurs, depends on the layer thickness and on its structural quality.

The most stable crystalline configuration of NbTiN is the rocksalt cubic structure, with a lattice parameter between $a_{NbN}$ = 4.39 Å [17] and $a_{TiN}$ = 4.24 Å [20]. The use of III-nitride semiconductors (GaN, AlN) as substrates for NbTiN is interesting because the equivalent lattice constant of NbN in the (111) plane $a_{NbN}^{(111)} = 3.104$ Å is closely lattice matched to the in-plane lattice constant of wurtzite AlN ($a_{AlN}$ = 3.112 Å, mismatch < 1%) and GaN ($a_{GaN}$ = 3.189 Å, mismatch ≈ 3.5%). These III-N materials present the additional interest of being optically transparent in a large spectral range, from their band gap (200 nm and 365 nm for AlN and GaN, respectively) to the far infrared, with the exception of the 9.6-19 µm and 7.3-9 µm regions that correspond to the first and second harmonics of the reststrahlen band [21,22].

It has been reported that the use of crystalline GaN, AlGaN or AlN as substrates results in an improvement of the crystalline quality and critical superconducting temperature of Nb(Ti)N thin films, which grow epitaxially on (0001)-oriented III-N [16–18,23]. However, there is almost no information on the performance of SNSPDs fabricated on III-nitrides. In a conference contribution, Zhu et al. showed NbN SNSPDs on crystalline AlN-on-sapphire with reduced jitter but poor detection efficiency (< 4%) [24]. On the other hand, the use of sputtered (polycrystalline) AlN as a buffer layer for the deposition of NbN on GaAs substrates resulted in improved performance of SNSPDs in comparison with detectors fabricated on bare GaAs [25]. However, the efficiency of such devices was only 5% at 900 nm. It increased up to 31% at 385 nm, but the substrate absorption at that wavelength would prevent the implementation of the microcavity or waveguide based designs required to get close-to-unity efficiencies [26].

In this paper, we study the impact of the substrate on the crystalline quality and superconducting critical temperature of ultrathin (11 nm) films of NbTiN deposited on almost-lattice-matched III-nitride semiconductors by reactive magnetron sputtering. Various materials are considered as substrates, namely a 1-µm-thick AlN-on-sapphire template, a GaN/AlN (650 nm/1 μm) heterostructure on sapphire, and 350-μm-thick GaN substrates, either non-intentionally doped (n-type conductive) or Fe-doped (semi-insulating).

## 2. Methods

Thin (≈ 11 nm) NbTiN films were deposited by reactive magnetron co-sputtering at room temperature in an ultra-high vacuum chamber (background pressure between $5\times10^{-9}$ Torr and $1\times10^{-8}$ Torr). After solvent cleaning (acetone, methanol), the substrates were placed on a 4" wafer chuck and held in place using aluminum pins. During the deposition, the substrate holder was rotating in order to improve the radial uniformity of the samples. The substrates were held at a distance of about 30 cm from the targets. The Nb target was placed directly below the substrate holder, while the Ti target was placed in a confocal geometry, the target pointing towards the substrate holder. Both targets were 3" in diameter and 0.250" thick. The applied power on the Nb target was 200 W (direct current), while the applied power on the Ti target was 200 W (radiofrequency), which corresponds to a power density of 4.39 W/cm$^2$. The sputtering pressure was set to 3 mTorr with a gas mixture of 100 sccm of Ar and 10 sccm of $N_2$. A quartz crystal monitor was calibrated by deposition of a thick film and subsequent measurement of the deposited thickness with a profilometer. From measurements on previous samples deposited in the same conditions, we estimate the thickness variation in 1 cm × 1 cm, to be less than 1%. The estimated deposition rate was 0.1 nm/s. The samples were deposited two at a time during subsequent runs, and under similar conditions, namely same pressure, gas partial pressures, applied powers, plasma voltage and current. The nominal Ti mole fraction in the $Nb_{1-x}Ti_xN$ layer was x = 0.2, but chemical measurements by electron energy loss spectroscopy and energy-dispersive x-ray spectroscopy of reference specimens, performed in a transmission electron microscope, point to x = 0.15±0.03. Substrates consisted of

(i) a commercial template (AlN-on-sapphire template) consisting of 1-µm-thick (0001)-oriented AlN deposited by metalorganic vapor phase epitaxy (MOVPE) on c-plane sapphire (sample SA),

(ii) a 650-nm-thick (0001)-oriented GaN layer deposited by plasma-assisted molecular-beam epitaxy (PAMBE) on a commercial template consisting of 1-µm-thick AlN deposited by MOVPE on c-plane sapphire (sample SAG),

(iii) a 350-µm-thick (0001)-oriented GaN substrate (sample SG) diced from thick GaN boules grown by hydride vapor phase epitaxy on c-plane sapphire, and

(iv) a 350-µm-thick (0001)-oriented semi-insulating Fe-doped GaN substrate (sample SGF) diced from thick Fe-doped GaN boules grown by hydride vapor phase epitaxy on c-plane sapphire.

Prior to the deposition, the surface was treated *in situ* with a 25 W RF Ar plasma for 10 s. The samples under study are listed in table 1.

X-ray reflectivity and x-ray diffraction measurements (XRD) were performed in a Panalytical Empyrean diffractometer equipped with a cobalt anode beam tube (Co $\lambda_{K\alpha}$ = 1.789 Å), a Göbel mirror and a two-dimensional pixel detector used in both zero-dimensional and one-dimensional modes.

The microstructure of the samples was analyzed by high-resolution transmission electron microscopy (HRTEM). The samples were prepared in cross-section by either mechanical polishing followed by ion milling or by focused ion beam milling and measured in a FEI-Tecnai microscope operated at 200kV. Finally, the surface morphology was measured by atomic force microscopy (AFM) in a Dimension 3100 system operated in the tapping mode.

## 3. Results and discussion

### 3.1 Material characterization

The thickness of the NbTiN layer has been determined precisely using x-ray reflectivity. Examples of these measurements corresponding to samples SA and SGF are illustrated in figure 1, together with fits using the GenX software [27]. For all the samples, the fit confirms total layer thickness around 11 nm (see precise values and error bars in table 1). Getting a good agreement with the experimental results requires assuming a multilayer structure consisting of a bottom $Nb_{0.8}Ti_{0.2}N$ layer ($\approx$ 8.45±0.45 nm) capped by an $Nb_{0.8}Ti_{0.2}N_{0.9}O_{0.1}$ layer ($\approx$ 2.1±0.2 nm) and by a mixture of $Nb_2O_5$ and $TiO_2$ [less than one atomic layer, with $(Nb_2O_5)_{0.7}(TiO_2)_{0.3}$ composition]. Post-growth oxidation and $Nb_2O_5$ and $TiO_2$ demixing are known to occur in NbTiN thin films, but experiments as a function of time show a saturation so that it remains a surface phenomenon [28]. Furthermore, due to the hydrophilic nature of the material, a further improvement of the fit is observed when assuming the presence of a layer of $H_2O$ at the surface (see figure 1). The values of root-mean-square (RMS) roughness of the topmost layer extracted from the calculations are also summarized in table 1, considering the fits with and without water termination.

To identify the crystalline orientation of the NbTiN film and its epitaxial relationship with the substrates, structural studies were performed by HRTEM along both [11-20] and [10-10] zone axis of the III-nitride substrates. Figure 2(a) show sample SAG viewed along the [11-20] zone axis of GaN. The superconducting film grows along the [111] axis, and it contains twin domains that are rotated by 60°. The images confirm the nominal crystallographic orientation of the substrates, and the establishment of

NbTiN(111) [2–1–1] || (Al,Ga)N (0001) [10–10]  and

NbTiN(111) [2–1–1] || (Al,Ga)N(0001) [01-10]  epitaxial relationships, as previously observed in the case of NbN layers [10,18,23]. The presence of dislocations at the interface is consistent with the relaxation of the misfit (36 GaN planes correspond to 37 NbTiN planes, as illustrated in figure 2(b)). In the case of SGF, observed in the same orientation, the thin film presents the same structural characteristics, but a strong contrast is observed at the NbTiN/GaN:Fe interface (arrow in figure 2(c)). This contrast has been assigned to the segregation of Fe [29,30], which is used as a dopant to render the GaN insulating. Finally, figure 2(d) depicts the NbTiN/AlN interface in sample SA, viewed along the [10-10] zone axis of AlN. The interface presents an epitaxial relationship as observed for NbTiN on GaN, but with lower density of dislocations, which appear separated by more than 100 AlN m-planes. This is consistent with the NbTiN lattice being more closely matched to AlN than to GaN.

The variation of the resistance with temperature for the samples under study is presented in figure 3. The superconducting critical temperature moves in the range from $T_c$ = 10.6 to 11.8 K, depending on the substrate, with a superconducting transition range (over which the resistance varies from 90% to 10% of its maximum value) $\Delta T$ = 0.83-0.86 K. Precise values of $T_c$ and $\Delta T$ for the samples under study are summarized in table 1, and they represent a significant improvement in comparison with our typical results of similar NbTiN layers deposited on dry thermal $SiO_2$ on Si, whose $T_c$ is in the range of 9.8±0.3 K with $\Delta T \approx 0.87$. A similar enhancement of $T_c$ was observed by Shiino *et al.* when inserting a sputtered AlN buffer layer between the NbTiN thin film and the quartz substrate [17]. This improvement is attributed to the good lattice matching between (111)-oriented rock-salt NbTiN and (0001)-oriented wurtzite AlN.

The superconducting properties of NbTiN might depend on its strain state [31]. In this sense, Makise et al. showed an increase of $T_c$ in NbTiN ultrathin films with a moderate tensile strain (out-of-the-plane lattice constant $a_\perp$ around 4.35-4.39 Å in layers with a relatively high Ti content, around 30%) grown on various substrates. Krause *et al.* report a degradation of the superconducting properties of NbN when increasing the Al content of AlGaN substrates from GaN towards AlN [17], which implies reducing the lattice mismatch with the substrate. However, the possible correlation of this trend with the strain in the layers was not analyzed. In our case, if we look at the in-plane lattice parameter of the substrate listed in table 1 (results extracted from x-ray diffraction measurements of the (0002) reflection of AlN or GaN, assuming biaxial strain), we see that it evolves from $a_{AlN}$ (SA) to $a_{GaN}$ (SG, SGF), passing through an intermediate state with $a$ = 3.170 Å due to the partial relaxation of the thin GaN layer on AlN for sample SAG. This slow relaxation is characteristic of III-nitride films grown by PAMBE under metal-rich conditions [32]. If we compare the evolution of $T_c$ with the lattice parameter of the substrate (both listed in table 1), we observe an increase of $T_c$ for decreasing substrate lattice parameter, i.e. a systematic improvement of the superconducting properties as we reduce the lattice mismatch between NbTiN and the substrate.

To get a precise assessment of the strain state of the superconducting layer, a detailed XRD study was conducted. The (111) inter-planar distance of the NbTiN layers along the [111] growth axis ($d_{[111]}$) was extracted from the angular position of the (111) and (222) reflections in ω–2θ scans, whereas the inter-planar distance along the in-plane [2–1–1] axis ($d_{[2-1-1]}$) was extracted from the angular position of the (311), (133), (224), and (511) reflections. Average values of $d_{[111]}$ and $d_{[2-1-1]}$ are presented in table 1. From such values, we can calculate the equivalent out-of-the-plane and in-plane lattice parameters, $a_\perp$ =

$\sqrt{3}\ d_{[111]}$ and $a_{\parallel} = \sqrt{6}\ d_{[2–1–1]}$, respectively, also presented in the table. For all the samples, $a_{\perp}$ and $a_{\parallel}$ are identical, within the error bars of the measurements, which means that the samples are fully relaxed.

Figure 4 compares the surface morphology of the NbTiN layers under study, measured by atomic force microscopy in the tapping mode. Typical AFM images shown in figures 4(a) and (b) (describing samples SA and SGF, respectively) show thin grainy layers. The surface roughness presents significant differences from sample to sample, as summarized in figure 4(c). Comparing samples SA and SGF, the average height increases from 1.9 to 4.7 nm, while the RMS roughness evolves from 0.85 to 1.6 nm, respectively. Note that these values are higher than the surface roughness of the substrates (shown in table 1), and there is no correlation between the NbTiN roughness and the substrate roughness. On the contrary, the variation of the NbTiN roughness from sample to sample correlates well with the lattice mismatch between the superconducting film and the substrate, which confirms that the strain relaxation plays a role in the surface roughness development. Let us remind that for low misfits and ultra-thin layers, relaxation can be achieved by elastic undulation of the surface prior to the formation of dislocations [33].

From the structural analysis of the NbTiN layers we conclude that there is an enhancement of $T_c$ in epitaxial layers grown on III-nitrides with respect to layers deposited on amorphous materials, like $SiO_2$. A further improvement is observed as the substrate gets more closely lattice matched to the NbTiN epilayer, i.e. the $T_c$ is higher for material deposited on AlN with respect to deposition on GaN. This trend cannot be related to the strain state, since the samples are fully relaxed. We therefore attribute it to the reduced roughness of the NbTiN film. This result might also explain the results observed by Krause et al. when depositing NbN on various $Al_xGa_{1-x}N$, buffer layers [23], who showed a decrease of $T_c$ for increasing Al mole fraction. In absence of studies of the strain state of the superconductor, an enhancement of the roughness in the ternary alloys could explain the degradation of the superconducting performance.

### 3.2 Fabrication and characterization of SNSPDs

In order to assess the suitability of such NbTiN films to the foreseen application, we have fabricated SNSPDs from sample SA, which displayed the best superconducting properties. The devices were designed for operation under normal incidence illumination. The NbTiN film was patterned into a meander structure as described in ref. [8], using e-beam lithography and inductively-coupled plasma etching. The diameter of the detector area was 15 µm. The nanowire width was 70 nm, with a pitch of 150 nm, as illustrated by the scanning electron microscopy (SEM) image in the insert of figure 5(a).

We used a dip-stick system to immerse the detector in liquid helium. In our system, the detectors and the internal electronics are immersed within liquid helium, ensuring a stable operation temperature of 4.2 K. Continuous-wave laser diodes emitting at wavelengths λ = 1.31 µm and 0.65 µm were used for optical excitation. The laser power was attenuated in order to obtain a photon flux around $10^6$ photon/s. Figures 5(a) and 5(b) show the evolution of the detection efficiency at λ = 1.31 µm and 0.65 µm, respectively, as a function of the bias current/critical current ratio ($I_b/I_{cr}$). The dark count rate is superimposed on figure 5(a). As expected [34], the efficiency is maximum for transverse-electric (TE) polarization, where the electric field oscillates along the nanowire length, and drops down for transverse-magnetic (TM) polarized light, where the electric field oscillates along the nanowire width. At λ = 1.31 µm, the highest efficiency recorded was 21% for TE polarization, but the response is not fully saturated. On the contrary, at λ = 0.65 μm, the higher photon energy results in full saturation for $I_b/I_{cr} \approx 0.8$ at 4.2 K.

Note that the devices present a very low level of dark counts, about 10 times lower than typical measurements in SNSPDs on $SiO_2$/Si substrates. The low dark count rate is explained by the high critical current (35±3 μA, which corresponds to a critical current density in the meander of ≈ 5.9 MA/cm$^2$). Therefore, the voltage trigger level is well above the noise level of the readout system, which reduces the probability of dark counts. The enhancement of the critical current (and reduction of the noise level) is consistent with the improvement observed by Schmidth et al. when inserting a sputtered AlN buffer layer before the deposition of NbN on GaAs [25].

The device performance is modeled using finite-difference time-domain (FDTD) calculations with the commercial RSoft FullWave software [26,34]. The meander is modeled as a single grating period with in-plane boundary periodic conditions, as described in the inset of figure 5(c). Calculations assume an 8.5-nm-thick NbTiN active wire with a width of 70 nm and a pitch of 150 nm. The absorption efficiency is calculated from the difference between the input power and the reflected plus transmitted power, once the steady state is reached. The refractive indices used as input parameters were 4.17 + i5.63 and 2.08 + i2.95 for NbTiN at 1.31 μm and 0.65 μm, respectively, 1.74 for sapphire, and 2.15 for AlN.

Figures 5(c) and 5(d) present the simulated absorption efficiency of the NbTiN wire at λ = 1.31 μm and 0.65 μm, for TE and TM polarization, as a function of the thickness of the AlN template. The oscillatory patterns are due to Fabry-Perot interference in the AlN layer. The experimental values are represented with error bars that illustrate the uncertainty on the thickness of the AlN layer and the dispersion of efficiency values for several devices. At λ = 1.31 µm, the measurements approach the theoretical results, in spite of the fact that the experimental efficiency curves are not fully saturated. On the contrary, at λ = 0.65 µm, full saturation is reached, leading to an excellent fit between the experimental results and the simulations, which implies that the internal quantum efficiency is close to one. Therefore, these experiments can be considered as a validation of the approach from material point of view. Then, an enhancement of the external efficiency can be easily obtained by modifying the nitride layer thickness to create a microcavity operating at the desired wavelength, or by implementing a GaN waveguide, if the aim is broadband operation (note that the SAG sample has already the adequate layer sequence for waveguide fabrication) [26].

## 4. Conclusion

In conclusion, we studied superconducting and structural properties of ultrathin layers of NbTiN sputtered on III-nitride semiconductor substrates. The NbTiN films grow along the [111] axis, and they keep an epitaxial relation with the wurtzite substrate. The superconducting critical temperature is in the range of $T_c$ = 10.6-11.8 K, depending on the substrate, with a sharp superconducting transition ($\Delta T$ = 0.83-0.86 K). This represents an improvement with respect to layers deposited on amorphous substrates, like dry thermal SiO$_2$ ($T_c$ = 9.8±0.3 K, $\Delta T \approx$ 0.87). Higher $T_c$ with smaller $\Delta T$ is obtained for deposition on AlN, whose crystalline lattice is better matched to NbTiN than that of GaN. The improvement with respect to GaN-based substrates is not related to the strain state, since all the NbTiN layers are fully relaxed. Instead, we attribute the higher $T_c$ to a reduction of the NbTiN roughness when deposited on a lattice matched substrate. Based on these results, we have demonstrated the first superconducting nanowire single photon detectors on crystalline AlN, showing external quantum efficiencies that fit well with theoretical calculations.

## Acknowledgements

The authors acknowledge financial support from the French National Research Agency via the "WASI" (ANR-14-CE26-0007) program, and from the Grenoble Nanoscience Foundation via the "NAQUOP" project.

# References

[1] Shurakov A, Lobanov Y and Goltsman G 2016 Superconducting hot-electron bolometer: from the discovery of hot-electron phenomena to practical applications *Supercond. Sci. Technol.* **29** 023001

[2] Gol'tsman G N and Loudkov D N 2003 Terahertz Superconducting Hot-Electron Bolometer Mixers and Their Application in Radio Astronomy *Radiophys. Quantum Electron.* **46** 604–17

[3] Klapwijk T M and Semenov A V 2017 Engineering Physics of Superconducting Hot-Electron Bolometer Mixers *IEEE Trans. Terahertz Sci. Technol.* **7** 627–48

[4] Jiang L, Shiba S, Shiino T, Shimbo K, Sakai N, Yamakura T, Irimajiri Y, Ananthasubramanian P G, Maezawa H and Yamamoto S 2010 Development of 1.5 THz waveguide NbTiN superconducting hot electron bolometer mixers *Supercond. Sci. Technol.* **23** 045025

[5] Schuck C, Pernice W H P and Tang H X 2013 Waveguide integrated low noise NbTiN nanowire single-photon detectors with milli-Hz dark count rate *Sci. Rep.* **3** 1893

[6] Dauler E A, Grein M E, Kerman A J, Marsili F, Miki S, Nam S W, Shaw M D, Terai H, Verma V B and Yamashita T 2014 Review of superconducting nanowire single-photon detector system design options and demonstrated performance *Opt. Eng.* **53** 081907

[7] Tanner M G, Natarajan C M, Pottapenjara V K, O'Connor J A, Warburton R J, Hadfield R H, Baek B, Nam S, Dorenbos S N, Ureña E B, Zijlstra T, Klapwijk T M and Zwiller V 2010 Enhanced telecom wavelength single-photon detection with NbTiN superconducting nanowires on oxidized silicon *Appl. Phys. Lett.* **96** 221109

[8] Mukhtarova A, Redaelli L, Hazra D, Machhadani H, Lequien S, Hofheinz M, Thomassin J-L, Gustavo F, Zichi J, Zwiller V, Monroy E and Gérard J-M 2018 Polarization-insensitive fiber-coupled superconducting-nanowire single photon detector using a high-index dielectric capping layer *Opt. Express* **26** 17697

[9] Pernice W H P, Schuck C, Minaeva O, Li M, Goltsman G N, Sergienko A V and Tang H X 2012 High-speed and high-efficiency travelling wave single-photon detectors embedded in nanophotonic circuits *Nat. Commun.* **3** 1325

[10] Yan R, Khalsa G, Vishwanath S, Han Y, Wright J, Rouvimov S, Katzer D S, Nepal N, Downey B P, Muller D A, Xing H G, Meyer D J and Jena D 2018 GaN/NbN epitaxial semiconductor/superconductor heterostructures *Nature* **555** 183

[11] Mourik V, Zuo K, Frolov S M, Plissard S R, Bakkers E P A M and Kouwenhoven L P 2012 Signatures of Majorana Fermions in Hybrid Superconductor-Semiconductor Nanowire Devices *Science* **336** 1003–7

[12] Thouless D J, Kohmoto M, Nightingale M P and den Nijs M 1982 Quantized Hall Conductance in a Two-Dimensional Periodic Potential *Phys. Rev. Lett.* **49** 405–8

[13] Yang X, You L, Zhang L, Lv C, Li H, Liu X, Zhou H and Wang Z 2018 Comparison of Superconducting Nanowire Single-Photon Detectors Made of NbTiN and NbN Thin Films *IEEE Trans. Appl. Supercond.* **28** 1–6

[14] Merkel H F, Khosropanah P, Cherednichenko S and Kollberg E 2003 Comparison of the Noise Performance of NbTiN and NbN Hot Electron Bolometer heterodyne mixers at THz Frequencies *Proc. of the 14th International Symposium on Space Terahertz Technology* (Tucson, Arizona, US.) pp 31–2

[15] Miki S, Takeda M, Fujiwara M, Sasaki M, Otomo A and Wang Z 2009 Superconducting NbTiN Nanowire Single Photon Detectors with Low Kinetic Inductance *Appl. Phys. Express* **2** 075002

[16] Mercier F, Coindeau S, Lay S, Crisci A, Benz M, Encinas T, Boichot R, Mantoux A, Jimenez C, Weiss F and Blanquet E 2014 Niobium nitride thin films deposited by high temperature chemical vapor deposition *Surf. Coat. Technol.* **260** 126–32

[17] Shiino T, Shiba S, Sakai N, Yamakura T, Jiang L, Uzawa Y, Maezawa H and Yamamoto S 2010 Improvement of the critical temperature of superconducting NbTiN and NbN thin films using the AlN buffer layer *Supercond. Sci. Technol.* **23** 045004

[18] Sam-Giao D, Pouget S, Bougerol C, Monroy E, Grimm A, Jebari S, Hofheinz M, Gérard J-M and Zwiller V 2014 High-quality NbN nanofilms on a GaN/AlN heterostructure *AIP Adv.* **4** 107123

[19] Villegier J-C, Bouat S, Cavalier P, Setzu R, Espiau de Lamaestre R, Jorel C, Odier P, Guillet B, Mechin L, Chauvat M P and Ruterana P 2009 Epitaxial Growth of Sputtered Ultra-Thin NbN Layers and Junctions on Sapphire *IEEE Trans. Appl. Supercond.* **19** 3375–8

[20] Wriedt H A and Murray J L 1987 The N-Ti (Nitrogen-Titanium) system *Bull. Alloy Phase Diagr.* **8** 378–88

[21] Yang J, Brown G J, Dutta M and Stroscio M A 2005 Photon absorption in the Restrahlen band of thin films of GaN and AlN: Two phonon effects *J. Appl. Phys.* **98** 043517

[22] Welna M, Kudrawiec R, Motyka M, Kucharski R, Zając M, Rudziński M, Misiewicz J, Doradziński R and Dwiliński R 2012 Transparency of GaN substrates in the mid-infrared spectral range *Cryst. Res. Technol.* **47** 347–50

[23] Krause S, Meledin D, Desmaris V, Pavolotsky A, Belitsky V, Rudziński M and Pippel E 2014 Epitaxial growth of ultra-thin NbN films on Al$_x$Ga$_{1-x}$N buffer-layers *Supercond. Sci. Technol.* **27** 065009

[24] Zhu D, Choi H, Lu T-J, Zhao Q, Dane A, Najafi F, Englund D R and Berggren K K 2016 "Superconducting nanowire single-photon detector on aluminum nitride," 2016 Conference on Lasers and Electro-Optics (CLEO), 5-10 June 2016, San Jose, CA.

[25] Schmidt E, Ilin K and Siegel M 2017 AlN-Buffered Superconducting NbN Nanowire Single-Photon Detector on GaAs *IEEE Trans. Appl. Supercond.* **27** 1–5

[26] Redaelli L, Bulgarini G, Dobrovolskiy S, Dorenbos S N, Zwiller V, Monroy E and Gérard J M 2016 Design of broadband high-efficiency superconducting-nanowire single photon detectors *Supercond. Sci. Technol.* **29** 065016

[27] Björck M and Andersson G 2007 GenX: an extensible X-ray reflectivity refinement program utilizing differential evolution *J. Appl. Crystallogr.* **40** 1174–8

[28] Zhang L, You L, Ying L, Peng W and Wang Z 2018 Characterization of surface oxidation layers on ultrathin NbTiN films *Phys. C Supercond. Its Appl.* **545** 1–4

[29] Ishiguro T, Yamada A, Kotani J, Nakamura N, Kikkawa T, Watanabe K and Imanishi K 2013 New Model of Fe Diffusion in Highly Resistive Fe-Doped Buffer Layer for GaN High-Electron-Mobility Transistor *Jpn. J. Appl. Phys.* **52** 08JB17

[30] Balmer R S, Soley D E J, Simons A J, Mace J D, Koker L, Jackson P O, Wallis D J, Uren M J and Martin T 2006 On the incorporation mechanism of Fe in GaN grown by metal-organic vapour phase epitaxy *Phys. Status Solidi C* **3** 1429–34

[31] Makise K, Terai H, Takeda M, Uzawa Y and Wang Z 2011 Characterization of NbTiN Thin Films Deposited on Various Substrates *IEEE Trans. Appl. Supercond.* **21** 139–42

[32] Bellet-Amalric E, Adelmann C, Sarigiannidou E, Rouvière J L, Feuillet G, Monroy E and Daudin B 2004 Plastic strain relaxation of nitride heterostructures *J. Appl. Phys.* **95** 1127–33

[33] Perovic D D, Bahierathan B, Houghton D C, Lafontaine H and Baribeau J-M 1995 Strain Relaxation at Low Misfits: Dislocation Injection vs. Surface Roughening *MRS Proc.* **399**

[34] Redaelli L, Zwiller V, Monroy E and Gérard J M 2017 Design of polarization-insensitive superconducting single photon detectors with high-index dielectrics *Supercond. Sci. Technol.* **30** 035005

**Table 1.** Structural and electrical properties of the samples under study: nature of the substrate; substrate lattice parameter ($a_{Subs}$); thickness of NbTiN and total thin film thickness extracted from x-ray reflectivity (XRR), together with the RMS topmost surface roughness assumed for the fit, without and with water termination; out-of-the-plane and in-plane lattice parameters of NbTiN measured from XRD ($d_{[111]}$ and $d_{[2-1-1]}$, respectively) together with the cubic lattice parameters calculated from $d_{[111]}$ and $d_{[2-1-1]}$ assuming a regular cubic lattice ($a_\perp$ and $a_\parallel$, respectively); average peak height and RMS roughness extracted from AFM measurements of the NbTiN samples and of the substrates; from resistance vs. temperature measurements [R(T)], critical superconducting temperature ($T_c$) and temperature range of superconducting switching (ΔT) taken as the difference between the temperatures corresponding to a resistance drop of 10% and 90% from the maximum value.

|  | Sample | SA | SAG | SG | SGF |
|---|---|---|---|---|---|
|  | Substrate | Sapphire/AlN | Sapphire/AlN/GaN | GaN | GaN:Fe |
|  | $a_{Subs}$ (Å) | 3.112 | 3.170 | 3.189 | 3.189 |
| XRR | NbTiN thickness (nm) | 8.7±0.1 | 8.3±0.1 | 8.9±0.6 | 7.9±1.0 |
|  | Total thickness (nm) | 11.0±1.4 | 10.9±0.3 | 11.3±0.5 | 10.8±0.5 |
|  | RMS roughness (nm) | 3.5 | 1.5 | 0.8 | 2.0 |
|  | RMS roughness with H$_2$O (nm) | 0.5 | 0.6 | 1.1 | 1.8 |
| XRD | $d_{[111]}$ (Å) | 2.538±0.003 | 2.536±0.003 | 2.538±0.003 | 2.536±0.004 |
|  | $d_{[2-1-1]}$ (Å) | 1.797±0.010 | 1.791±0.005 | 1.798±0.004 | 1.791±0.005 |
|  | $a_\perp$ (Å) | 4.396±0.006 | 4.393±0.006 | 4.396±0.006 | 4.393±0.008 |
|  | $a_\parallel$ (Å) | 4.402±0.010 | 4.386±0.008 | 4.403±0.006 | 4.386±0.008 |
| AFM | av. peak height sample (nm) | 1.9 | 2.3 | 3.8 | 4.7 |
|  | RMS roughness sample (nm) | 0.85 | 0.85 | 1.5 | 1.6 |
|  | av. peak height substrate (nm) | 1.3 | 2.0 | 1.0 | 1.0 |
|  | RMS roughnes substrate (nm) | 0.15 | 0.5 | 0.11 | 0.11 |
| R(T) | $T_c$ (K) | 11.8 | 11.3 | 11.0 | 10.6 |
|  | $\Delta T$ (K) | 0.83 | 0.85 | 0.86 | 0.85 |

**Figure Captions**

**Figure 1.** X-ray reflectivity of (a) the NbTiN film on AlN-on-sapphire (sample SA), and (b) the NbTiN film on GaN:Fe (sample SGF). Grey squares are experimental measurements and the solid lines are fits generated with the GenX software. Note that in the case of SA, fits with and without $H_2O$ wetting layer are almost superimposed.

**Figure 2.** (a,b) HRTEM images of the NbTiN layer and the NbTiN/GaN interface of sample SAG viewed along the [11-20] zone axis of GaN. (c) HRTEM image of SGF viewed along the [11-20] zone axis of GaN. The arrow marks the contrast at the NbTiN/GaN:Fe interface. (d) HRTEM image of the NbTiN/AlN interface in sample SA, viewed along the [10-10] zone axis of AlN.

**Figure 3.** Variation of the resistance of the NbTiN layers under study as a function of temperature. The curve for an NbTiN deposited on dry thermal $SiO_2$ on Si under identical conditions was added for comparison. The resistance of the various samples was normalized to its maximum value.

**Figure 4.** Atomic force microscopy images of samples (a) SA and (b) SGF. (c) Representation of the average height and RMS roughness extracted from 800×800 $nm^2$ AFM images of samples SA, SAG, SG and SGF.

**Figure 5.** (a) Detection efficiency as a function of the bias current / critical current ratio ($I_b/I_{cr}$), measured for an SNSPD device fabricated on sample SA and measured at 1.31 μm and at 4.2 K (liquid He immersion). The dark count rate is also represented. Inset: SEM image of a detail of the superconducting nanowire meander. (b) Detection efficiency as a function of $I_b/I_{cr}$, measured at 0.65 μm and at 4.2 K. The dark count rate is also represented. (c) Calculated dependence of TE and TM absorption efficiency at 1.31 μm as a function of the thickness of the AlN layer. Experimental results are presented as squares with their associated error bars. Inset: Scheme of the structure for which simulations have been performed. (d) Calculated dependence of TE and TM absorption efficiency at 0.65 µm as a function of the thickness of the AlN layer. Experimental results are presented as squares with their associated error bars.

**Figure 1**

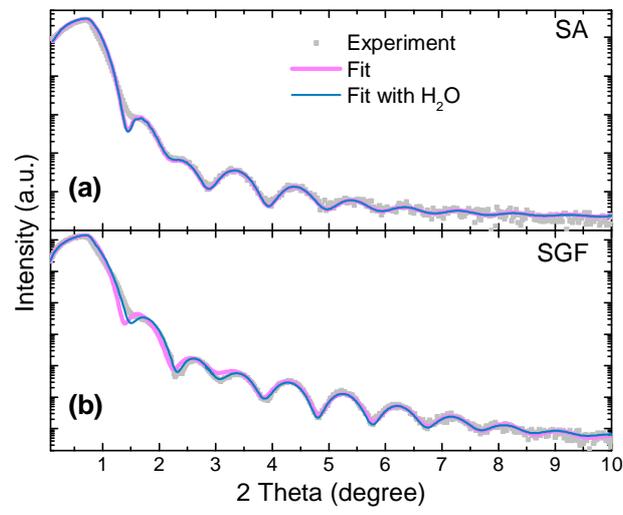

**Figure 2**

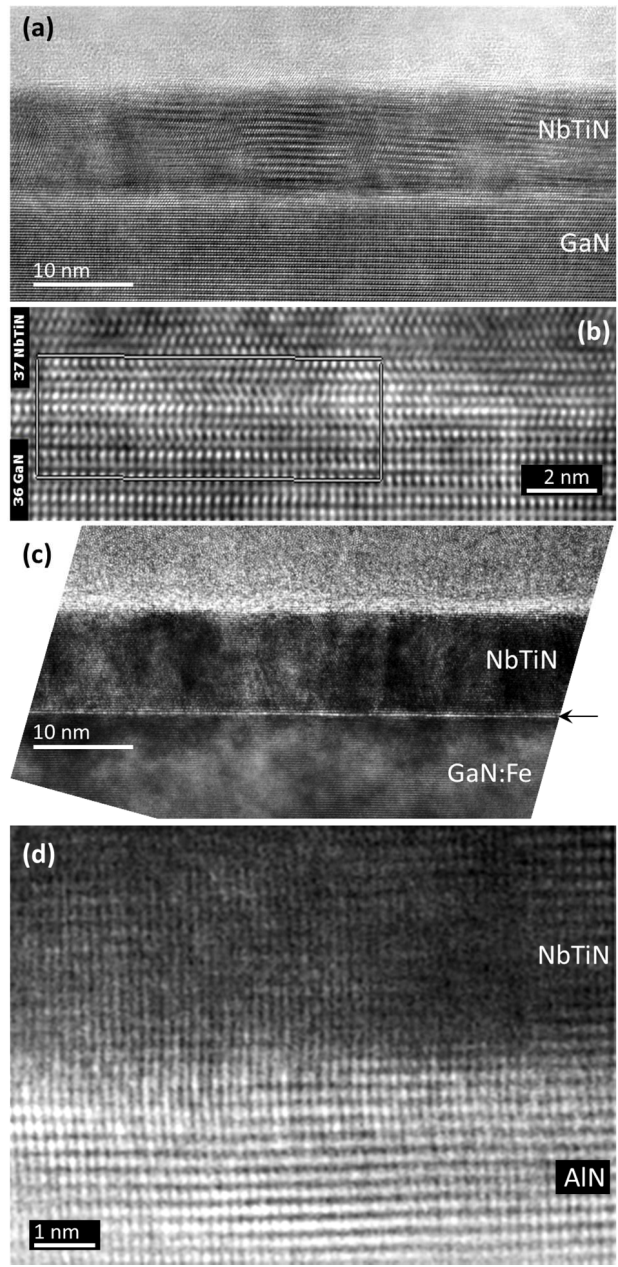

**Figure 3**

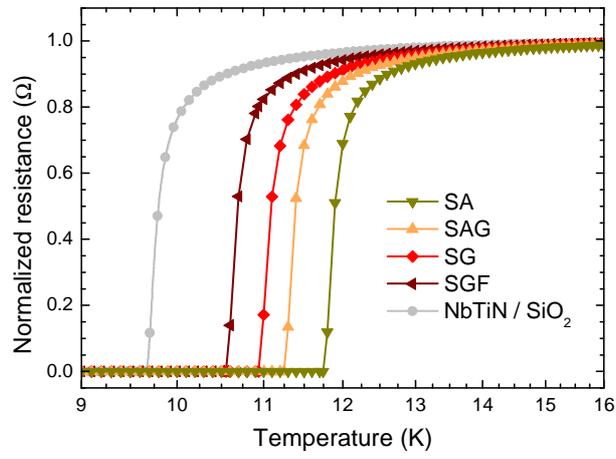

**Figure 4**

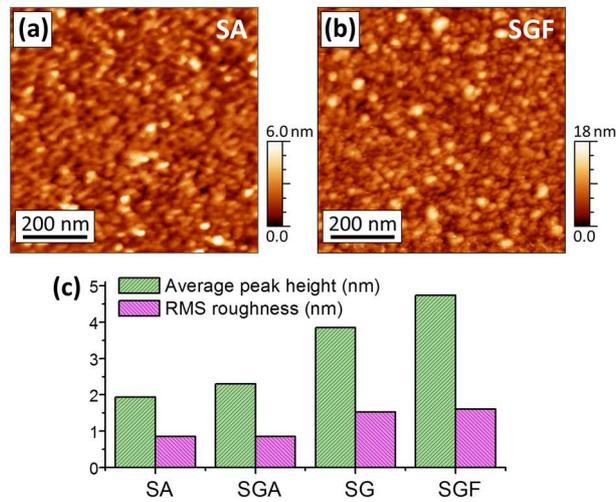

**Figure 5**

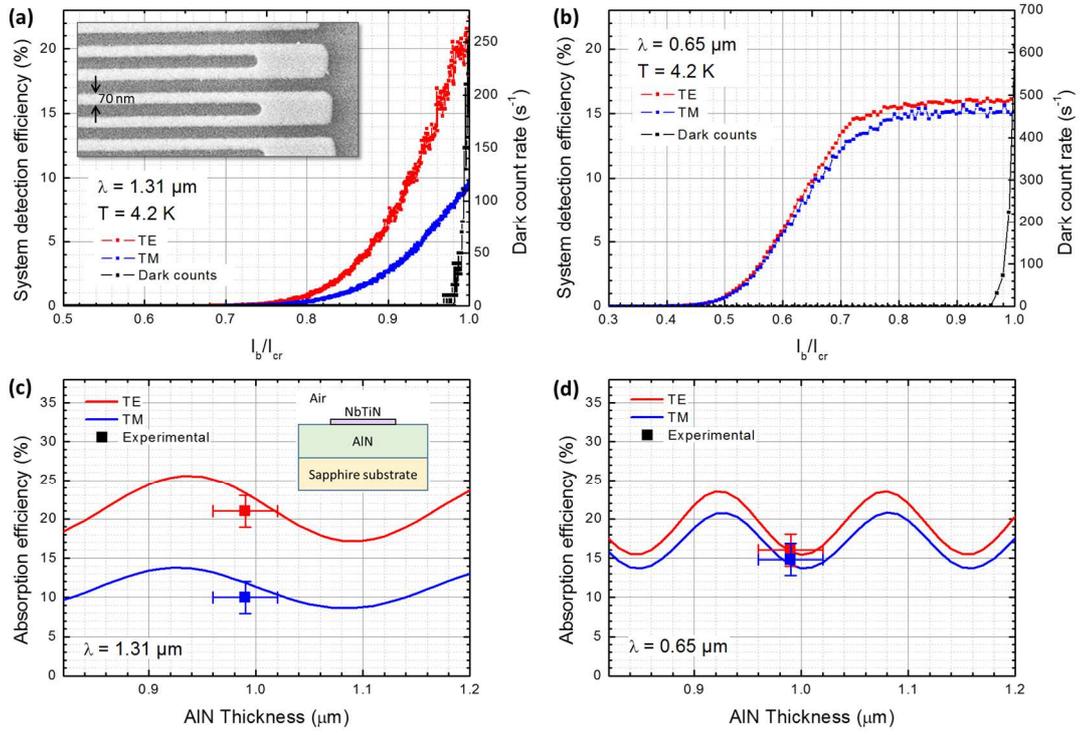